\documentclass[12pt]{book}
\usepackage{graphicx}

\begin{document}

\title{\bf eBook series:
Frontiers in Nuclear and  Particle Physics \\
Special Issue: Multielectronic Processes in Collisions Involving
Charged Particles and Photons with Atoms and Molecules \\
\vskip 0.3cm \large{in press since 26 June 2017}}

\frontmatter                            
\maketitle                              

\chapter[Chapter 5]{Chapter 5 \\ Multielectronic processes in particle and antiparticle collisions with rare gases}
\author{\large{Claudia C. Montanari}
\\ \vskip0.4cm
\noindent{\small{Instituto de Astronom\'{\i}a y F\'{\i}sica del
Espacio, Consejo Nacional de Investigaciones Cient\'{\i}ficas y
T\'ecnicas, and Universidad de Buenos Aires, casilla de correo 67,
sucursal 28, C1428EGA, Buenos Aires, Argentina.}} }

\vskip 0.3cm
\maketitle                              

\section*{Summary}In this chapter we analyze the multiple ionization
 by impact of $|Z|=1$ projectiles:
electrons, positrons, protons and antiprotons. Differences and
similarities among the cross sections by these four projectiles
allows us to have an insight on the physics involved. Mass and
charge effects, energy thresholds, and relative importance of
collisional and post-collisional processes are discussed. For this
purpose, we performed a detailed theoretical-experimental comparison
for single up to quintuple ionization of Ne, Ar, Kr and Xe by
particles and antiparticles. We include an extensive compilation of
the available data for the sixteen collisional systems, and the
theoretical cross sections by means of the continuum distorted wave
eikonal initial state approximation. We underline here that
post-collisional ionization is decisive to describe multiple
ionization by light projectiles, covering almost the whole energy
range, from threshold to high energies. The normalization of
positron and antiproton measurements to electron impact ones, the
lack of data in certain cases, and the future prospects are
presented and discussed.

\tableofcontents

\listoffigures

\mainmatter

\newpage
\section{Introduction}\label{s1}

Multiple ionization is a challenging subject, which plays an
important role in the knowledge of many-electron processes, such as
multiple-electron transitions, collisional and post-collisional
ionization or electron correlation effects. The goal of this
contribution is to deepen in the study of the multielectronic
processes by collision of $|Z|=1$ particles and antiparticles. For
this purpose we focused in the multiple ionization of the heaviest
rare gases, Ne, Ar, Kr and Xe. We analyzed the differences and
similarities in the cross sections of equal-charge versus equal-mass
projectiles. The ionization cross sections by light (electrons and
positrons) and heavy (protons and antiprotons) projectiles are quite
different in the low and intermediate energy regions. On the
contrary, in the high energy region all these cross sections
converge. However, this convergence to proton impact values is
different for antiprotons, positrons or electrons. A detailed
knowledge of this tendency from intermediate to high impact energies
is important because of the experimental normalization of relative
cross sections of antiparticles to electron or to proton values.
Classical reviews on particle and antiparticle collisions can be
found in \cite{McGuire88,Schultz91,knudsen,Knudsen92,P-L}.

The theoretical description of the multiple ionization processes by
these projectiles must consider the charge and mass effects, the
projectile trajectories, and the energy thresholds. The
post-collisional ionization (PCI) due to Auger-type processes
following inner-shell ionization, enhance the final number of
emitted electrons. For heavy projectiles this is important at high
energies. For light projectiles, such as electrons and positrons,
PCI dominates the highly-charged ion production in the whole energy
range, even close to the energy threshold. The ionization by proton
and electron impact has been studied since the early years of the
development of atomic physics. However, the experimental data on
multiple ionization by high energy protons could not be
theoretically described until the last fifteen years
\cite{Cavalcanti02,Cavalcanti03,Spranger04,Galassi07,MM1}; and by
electron impact only recently \cite{e-rare}. This was possible by a
consistent inclusion of the branching ration for PCI within the
independent electron model.

Multiple ionization is also a sensitive test for the experimental
work. Measurements require highly advanced techniques to get all
possible channels and final states. For protons, they must separate
pure ionization from capture channels, which enhance the data in the
intermediate energy region \cite{DuboisPRL84}. In the case of
positron impact, the total ionization values at low energies include
positronium formation \cite{Laricchia13}. On the other hand, the
higher the order of ionization the smaller the cross sections. The
quintuple ionization cross sections of Kr and Xr are of the order of
the $10^{-19}$ cm$^2$ at high energies, while for Ar they are
$10^{-20}$  cm$^2$ or even $10^{-21}$  cm$^2$.

The published experimental data of multiple ionization of Ne, Ar, Kr
and Xe by proton impact is profuse
\cite{DuboisPRL84,Cavalcanti02,Cavalcanti03,haugen,Dubois84,Andersen87,DuboisM,Manson87,%
gonzalez,sarkadi}. Much more abundant are the measurements by
electron impact. They include the pioneering works by Schram and
co-workers in the 60s \cite{schram} to the present
\cite{nagy,syage,krish,Rejoub,kobayashi,mccallion,straub,almeida,liebius,singh,koslowski}.

Instead, the experimental data by antiparticle is more scarce, what
is reasonable. Antiprotons are produced in high-energy physics
sources and then decelerated for atomic collisions experiments. This
research has been developed by Knudsen and coworkers at CERN
\cite{Andersen87,Andersen86,Andersen89,Paludan97,knudsenPRL08,knudsenNIMB09},
first in the low energy antiproton ring (LEAR) and nowadays in the
antiproton decelerator (AD). A very recent {\it state of art} of
antiproton impact ionization has been published by Kirchner and
Knudsen \cite{K-K}. In the case of positrons, most of the
experimental publications report only values of single ionization
\cite{knudsen90,Jacobsen95,Mori94,Laricchia02,Marler05}, some
articles include double or triple ionization measurements
\cite{Charlton89,Kara97,Moxom96,Moxom99,bluhme99,Moxom00,helms,kruse},
and only one paper quadruple ionization \cite{Moxom00}. It is worth
to mention that antiproton and positron values are normalized to
electron impact ionization cross sections at high energies.

In this chapter we present a comparison of the multiple ionization
cross sections by impact of electrons, positrons, protons and
antiprotons. We consider the rare gases Ne, Ar, Kr and Xe, and final
charge states from +1 to +3 (Ne), and +5 (Ar, Kr and Xe).  The
comparison includes the theoretical results obtained by employing
the continuum distorted wave eikonal initial state approximation in
\cite{MM1,e-rare,AP,positron}, and the available experimental
measurements. The extensive compilation of data for the four
projectiles and the four targets included here, allow us to have a
wide vision of the experimental \emph{state of art}, and some future
prospects.

\section{Theoretical description}\label{s2}

The theoretical description of the multiple ionization reviewed in
this chapter relies on three approximations:

\begin{enumerate}
\item The \textbf{independent particle model} (IPM): the ejected electrons ignore
each other, neglecting the correlation in the final state and the
changes in the target potential due to the successive loss of
electrons. Under this assumption, the probability of multiple
ionization can be expressed as a multinomial combination of
independent ionization probabilities.
\item The \textbf{continuum distorted wave eikonal initial state} (CDW-EIS)
approximation \cite{miraglia08,FPR}. This is a proved model to
describe intermediate and high energy multiple ionization by protons
and antiprotons \cite{MM1,AP}. In \cite{e-rare,positron}, the
CDW-EIS is adapted to describe ionization by electrons or positrons,
by taking into account the finite momentum transferred, the
non-linear trajectory and the mass effect. Light particle ionization
is characterized by the sharp energy thresholds, which are different
for single to quintuple ionization. The results in \cite{positron}
include the different thresholds of energy within the multinomial
expansion following \cite{hci}.
\item The \textbf{post-collisional ionization (PCI), independent of the projectile}. Thus, it is included within the multinomial expression in a semi
empirical way following \cite{Cavalcanti02,Spranger04}, using the
experimental branching ratios of the charge-state distribution after
a single initial vacancy. Present formalism is explained in
\cite{MM1}. A detailed compilation of these branching ratios is
available in \cite{e-rare}.
\end{enumerate}

In what follows the implications of these approximations on the
calculations are summarized. More details of them can be found in
the mentioned references.

\subsection{The independent particle model for multiple ionization}
\label{s2.1}

Within the IPM, the probability of direct ionization of exactly
$q_{j}$ electrons of the $j$ subshell as a function of the impact
parameter $b$, $P_{(q_{j})}(b)$, is obtained as a multinomial
distribution of the ionization probabilities $p_{j}(b)$ given by
\begin{equation}
P_{(q_{j})}(b)=\left(\matrix {N_{j} \cr q_{j}} \right) \
[p_{j}(b)]^{q_{j}}[1-p_{j}(b)]^{N_{j}-q_{j}}, \label{binomial}
\end{equation}
where  $N_{j}$ is the total number of electrons in the subshell. If
$n$ electrons are ionized from the different shells,
$n=\sum_{j}q_{j}$, then the total probability of direct ionization
is
\begin{equation}
P_{(n)} (b)=\sum_{q_{1}+q_{2}+...=n}\ \prod_{j}P_{(q_{j})}(b).
\label{multinom}
\end{equation}
and the cross section corresponding to the direct ionization of
exactly $n$ electrons is
\begin{equation}
\sigma_n=\int \ P_{(n)} (b) \ 2 \pi \ b\ db. \label{directCS}
\end{equation}

The main and more sensitive values of these calculations are the
ionization probabilities as function of the impact parameter,
$p_{j}(b)$. Different models have been employed in the last fifteen
years to obtain these $p_{j}(b)$ for multiple ionization
calculations, by the groups of Montenegro
\cite{Cavalcanti02,Cavalcanti03,MM1,Santos01,santana,B2,icpeac11},
Kirchner
\cite{Spranger04,K01,Kirchner05,KirchnerAr,Kirchner13,Kirchner15},
Rivarola \cite{Galassi07,tachino,tachino2}, and Miraglia
\cite{MM1,e-rare,AP,positron,sigaud13,archubi}.

\subsection{CDW-EIS ionization probabilities by proton, antiproton,
electron and positron impact} \label{s2.2}

We review here the results for proton, antiproton, electron and
positron impact multiple ionization in
\cite{MM1,e-rare,AP,positron}. These results employ the CDW-EIS code
by Miraglia \cite{miraglia08}. Details of these calculations are in
\cite{AP,miraglia08}. The aim is to discuss the scope and
limitations of this model to deal with particle-antiparticle and
heavy-light projectile effects.

The CDW-EIS ionization probabilities $p_{j}(b)$ as function of the
impact parameters are the seeds to be introduced in the multinomial
expression given by (\ref{binomial}). As expected, these results
tend to the first Born approximation ones at sufficiently high
impact energies \cite{AP}. In fact, the extension of the theoretical
results for impact velocities $v> 8$ a.u. in
\cite{MM1,e-rare,AP,positron} is achieved using the first Born
approximation.

On the other hand, the CDW-EIS results for light projectiles in
\cite{e-rare,positron} were obtained by adapting the CDW-EIS model
for equal-charged light projectiles, i.e. protons $\to$ positrons;
antiprotons $\to$ electrons. This calculation accounted for the
finite momentum transferred, the non-linear trajectory (very
different for positrons and electrons due to the repulsive or
attractive potentials), and the minimum energy for ionization. A
detailed explanation and the corresponding equations can be found in
\cite{e-rare}.

The CDW-EIS results in \cite{MM1,e-rare,AP,positron} describe
\emph{pure} ionization of the target. No electron transfer or
transfer followed by projectile electron loss is included. These
processes are possible only for positive projectiles, and may
enlarge the experimental cross sections at low energies.

Ionization processes present a sharp energy threshold: ionization is
not possible for projectile energies below the binding energy of the
target outer electrons. The threshold for each shell of electrons
must be taken into account within the multinomial expression
(\ref{binomial}). One of the differences between light and heavy
projectiles is that, on equal impact velocity, light projectiles
have impact energies of the order of the target binding energies or
even lower. This is the main cause of differences between ionization
cross sections by light and heavy projectiles at intermediate to low
energies.

Experimentally the threshold or appearance energy is well known and
measured in electron and positron-impact ionization. For $n$-fold
ionization the experiments indicate that this appearance energy is
much greater than $n$ times the binding energy of the outermost
electrons. This was analyzed in \cite{hci} considering not only the
energy gap, but also the mean velocity of the outgoing electrons in
a semi-classical way. This proposal was included in the multiple
ionization calculations by electron and positron impact in
\cite{positron}. Although the threshold itself is rather well
described, the theoretical model cannot be employed close to the
energy threshold. The CDW-EIS approximation fails at low impact
energies, i.e. when the energy loss by the projectile is comparable
to the impact energy. We will return to this later in this chapter.

\subsection{Auger type postcollisional ionization}
\label{s2.3}

The PCI is the rearrangement of the target atom after inner-shell
ionization. Different processes (shake off, Auger decay and
emission) are involved in the emission of one or more electrons long
after the collision. Following Cavalcanti \emph{et al}
\cite{Cavalcanti02} and Spranger and Kirchner \cite{Spranger04}, the
PCI is included inside the binomial equation (\ref{binomial}) in a
semi-empirical way by using experimental branching ratios of charge
state distribution after single photoionization. The deeper the
initial hole, the greater the number of electrons in PCI.

The branching ratios $F_{j,k}$ are the probabilities of loosing $k$
electrons in PCI after single ionization of the $j$ subshell. They
correspond to time delayed processes that only depend on the target.
A detailed compilation of the experimental branching ratios
$F_{j,k}$ for Ne, Ar, Kr and Xe available in the literature is
tabulated in \cite{e-rare}. It is interesting to note that they
verify the unitary condition, $\sum_{k=0}^{k_{max}}F_{j,k}=1$. In
\cite{MM1} the branching ratios are introduced in (\ref{binomial})
as follows
\begin{equation}
P_{(q_{j})}(b)=\left(\matrix {N_{j} \cr q_{j}} \right) \ [p_{j}(b)\
\sum_{k=0}^{k_{max}}F_{j,k}]^{q_{j}}[1-p_{j}(b)]^{N_{j}-q_{j}}.
\label{binomial_PCI}
\end{equation}
Afterwards, the addition of probabilities is arranged in order to
put together those terms that contribute to the same final number of
emitted electrons. New probabilities of exactly $n$ emitted
electrons are obtained, including direct ionization and PCI,
$P_{(n)}^{PCI}$ (see section 2.3 in \cite{MM1} for details). Thus,
the corresponding cross sections of $n$-fold ionization including
PCI are
\begin{equation}
\sigma_n^{PCI}=\int \ P_{(n)}^{PCI} (b)\  2 \pi \ b\ db.
\label{PCICS}
\end{equation}

\section[Results and data]{Results and data of particle and antiparticle ionization: charge and mass effects}

What follows is a revision of the theoretical CDW-EIS results for
multiple ionization of Ne, Ar, Kr and Xe by protons and antiprotons
\cite{AP}, electrons \cite{e-rare}, and positrons \cite{positron}.
This is achieved by comparing and analyzing the cross sections for
the four $|Z|=1$ projectiles together, and contrasting them with the
experimental data available in the literature. The comparison is
performed on equal impact velocities, and is plotted as a function
of the corresponding electron impact energy. The change to proton
impact energies is straightforward, just an $m_p/m_e$ factor (the
ratio of masses of heavy and light projectiles).

In the following sections the results for Ne, Ar, Kr and Xe are
pesented. We will note that the threshold for light particles is
quite well described by the theory. However, we will focus on the
description of the intermediate to high impact energy processes that
is the range of validity of the CDW-EIS.

\subsection{The multiple ionization of Ne}\label{s3}

In figures \ref{fig1Ne}-\ref{fig3Ne} we plotted together the CDW-EIS
results for proton, antiproton, electron, and positron impact
ionization \cite{AP,e-rare,positron}, for final Ne$^+$ to Ne$^{+3}$.
We do not pretend to extend the validity of the IPM for Ne (ten
bound electrons) beyond the triple ionization \cite{icpeac11}.
Changes in the target potential may be significant in this case. On
the contrary, for targets with larger number of bound electrons, the
IPM is expected to work.

\subsubsection{Ne single ionization}

\begin{figure}[!h]
\begin{center}
\includegraphics*[width=30pc]{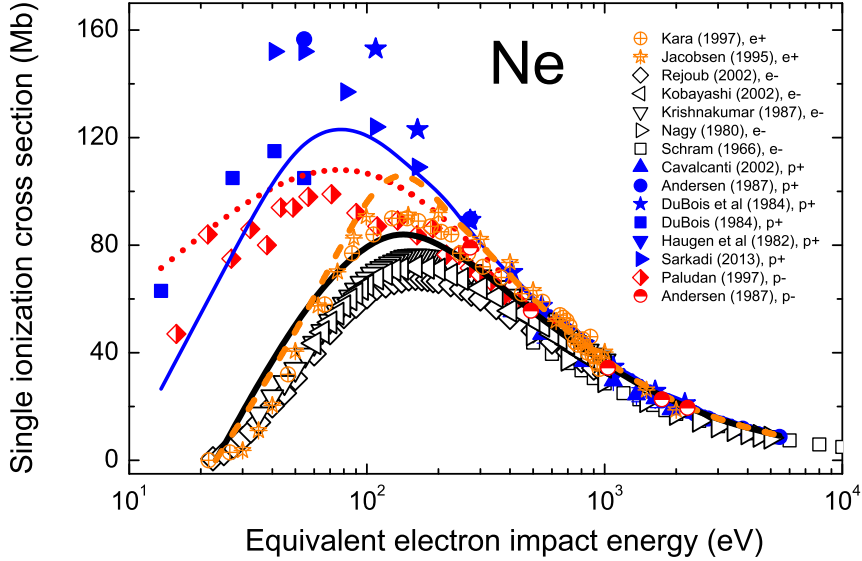}
\end{center}
\caption{\small{Single ionization of Ne by $|Z|=1$ projectiles as a
function of the impact energy, considering equal velocity for heavy
and light particles. Curves: CDW-EIS results for proton (blue thin
solid line), antiproton (red dotted line), electron (black thick
solid line) and positron (orange dashed-line) impact. Symbols:
details in the inset; the references are: for protons p+
\cite{DuboisPRL84,Cavalcanti02,haugen,Dubois84,Andersen87,sarkadi},
for electrons e- \cite{schram,nagy,krish,Rejoub,kobayashi}, for
positrons e+ \cite{Jacobsen95,Kara97}, and for antiprotons p-
\cite{Andersen87,Paludan97}.} \label{fig1Ne} }
\end{figure}

In figure \ref{fig1Ne} we include the theoretical curves and the
experimental measurements for the single ionization cross sections
of Ne by protons (p+), antiprotons (p-), electrons (e-), and
positrons (e+).
It can be observed that the main difference between heavy and light
projectiles are the lower values at intermediate and low energies,
and the sharp threshold around 23 eV. Instead, all the results
converge at high energies. The intermediate energy region is the
most interesting one. We note that for equal mass, the values for
projectiles with charge $Z=-1$ are below the $Z=+1$ ones. It could
be said that the ionization of Ne is more effective by positive
charges.

The theoretical description is rather good for the heavy
projectiles. In the case of protons, below the maximum the agreement
is fine only with the data by DuBois \cite{DuboisPRL84} who
separates pure ionization and capture. This may be the difference
between DuBois \cite{DuboisPRL84} data and the recent Sarkadi
{\textit et al} \cite{sarkadi} experimental values at intermediate
to low energies. In the case of antiprotons we remark the good
agreement with Andersen measurements at high energies
\cite{Andersen87}. Instead, the data by Paludan \cite{Paludan97} are
somewhat below the predictions, but the tendency is correct. For
single-ionization by light particles, the theoretical values
overestimate around the maximum, and describe nicely the
measurements at high energies.

The normalization of antiparticle cross sections is a point of
discussion \cite{positron}. The experimental cross sections by
positrons and antiprotons are mostly relative values normalized to
well known and tested high energy electron impact data, such as Rapp
\emph{et al} \cite{REG}, Sorokin \emph{et al} \cite{sorokinNe}, or
Krishnakumar {\textit et al} \cite{krish}. For example, the positron
data by Kara {\textit et al} \cite{Kara97} and antiproton data by
Paludan {\textit et al} \cite{Paludan97} are both normalized to
electron impact data by Krishnakumar {\textit et al} \cite{krish} at
800-1000 eV. The comparison displayed in figure \ref{fig1Ne} shows
that for the theoretical model, antiprotons and positrons tend
faster to proton values than to electron ones at high energies. At
400 eV proton, antiproton and positron curves are together and
$10\%$ above electron impact values. Even around 1000 eV the
electron single ionization is still a little below the rest. This
implies that antiproton and positron relative values could be
normalized to total proton-impact ionization cross sections, such as
those by Rudd et al \cite{Rudd}, at no so high energies.

\subsubsection{Ne double ionization}

\begin{figure}[h]
\begin{center}
\includegraphics*[width=30pc]{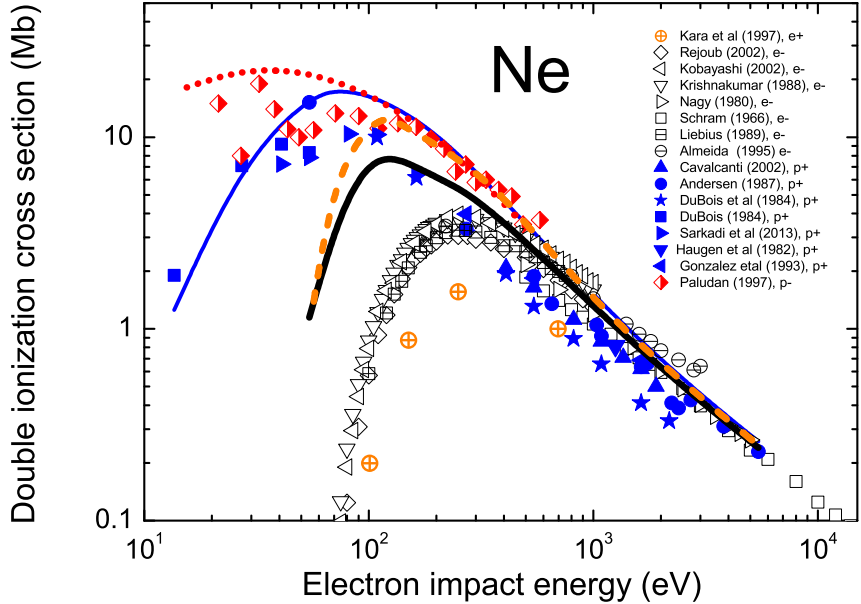}
\end{center}
\caption{\small{Double ionization of Ne by $|Z|=1$ projectiles as a
function of the impact energy, considering equal velocity for heavy
and light particles. Curves: CDW-EIS results for proton (blue thin
solid line), antiproton (red dotted line), electron (black thick
solid line) and positron (orange dashed-line) impact. Symbols:
details in the inset; the references are: for protons p+
\cite{DuboisPRL84,Cavalcanti02,haugen,Dubois84,Andersen87,gonzalez,sarkadi},
for electrons e-
\cite{schram,nagy,krish,Rejoub,kobayashi,almeida,liebius}, for
positrons e+ \cite{Kara97}, and for antiprotons p-
\cite{Paludan97}}.\label{fig2Ne} }
\end{figure}

In figure \ref{fig2Ne} we display the double ionization of Ne. We
use the logarithmic scale to emphasize the fall down and convergence
at high energies, which is much more drastic than for
single-ionization.

The double ionization cross sections by antiprotons at intermediate
to low velocities, i.e. $v<2.3$ a.u. are higher than by proton
impact. Inverse to single ionization. And this reversal can be seen
in the data and in the theoretical description too. Another point to
remark in figure \ref{fig2Ne} is that theoretically, above 200 eV
(electron impact) the proton, antiproton and even positron curves
converge to a single value. Instead, for electron impact the
convergence is above 600 eV. Experimentally, the positron data are
below the electron measurements. It is not clear how much it
influences the normalization of the single ionization measurements
of positrons to electron-impact data at the level of the double
ionization.

At high energies, the description is actually good. Our values
include the post-collisional Auger and shake off processes. At 3 keV
the PCI contribution is 60$\%$ of the total double ionization (see
figure 2 in \cite{positron}). At intermediate energies, the
theoretical curves clearly overestimate the electron and
positron-impact data. This is related to the inclusion of PCI due to
shake-off of the outer shell electrons of Ne, as explained in
\cite{e-rare}. This is an interesting open topic: The inclusion of
PCI allows us to describe the high energy region, but it produced
too high values at intermediate energies. On the contrary, if only
direct double ionization is considered, the maximum for electron
impact cross section and the energy threshold are better described,
but the high energy measurements are undervalued.

\subsubsection{Ne triple ionization}

\begin{figure}[h]
\begin{center}
\includegraphics*[width=30pc]{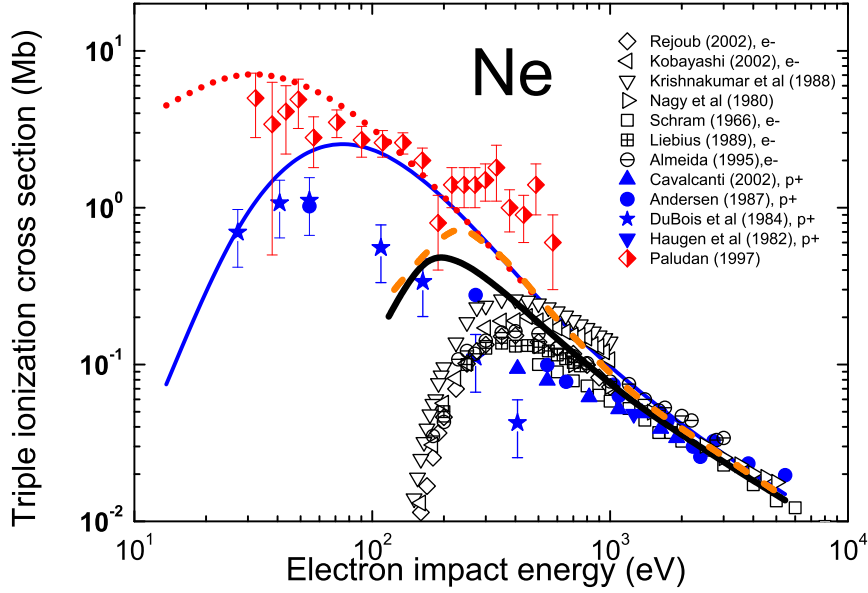}
\end{center}
\caption{\small{Triple ionization of Ne by $|Z|=1$ projectiles as a
function of the impact energy, considering equal velocity for heavy
and light particles. Curves: CDW-EIS results for proton (blue thin
solid line), antiproton (red dotted line), electron (black thick
solid line) and positron (orange dashed-line) impact. Symbols:
details in the inset; the references are: for protons p+
\cite{Cavalcanti02,haugen,Dubois84,Andersen87}, for electrons e-
\cite{schram,nagy,krish,Rejoub,kobayashi,almeida,liebius}, and for
antiprotons p- \cite{Paludan97}}.\label{fig3Ne} }
\end{figure}

The triple ionization cross sections of Ne are displayed in figure
\ref{fig3Ne}. The theoretical results for proton and antiproton
triple ionization are rather good. They clearly show that
antiprotons produce larger triple ionization for energies around the
maximum than protons. Note that this is exactly the opposite for
single ionization.

The comparison of the cross sections by heavy and light projectiles
shows the expected lower values for light projectiles and the
convergence at high energies. The CDW-EIS results for light
projectiles overestimate the measurements near the energy threshold.
This is partially due to the overestimation of the shake-off
contribution, and partially due to being in the limit of validity of
the model, as mentioned previously in this chapter.

It is very interesting to mention that the high energy data could
not be described if the postcollisional contribution to triple
ionization (initial ionization of inner shells and rearrangements
processes) is not included \cite{e-rare}. This contribution is
important above 400 eV, being $90\%$ of the triple ionization cross
sections at 1 keV.

\subsection{The multiple ionization of Ar}\label{s3}

We analyze in this section the Ar multiple ionization, from single
to quintuple. In figures \ref{fig1Ar}-\ref{fig5Ar} we displayed
together the CDW-EIS results for proton, antiproton, electron and
positron in Ar \cite{AP,e-rare,positron}, and an updated compilation
of experimental data for the four projectiles.

In the following figures it will be noted that the theoretical
description shows better agreement with the measurements than for
Ne. This behavior is enhanced for heavier targets, as we will
observe for Kr and Xe in the following sections. The IPM works
better to describe the multiple ionization in multielectronic
targets, so that the number of loose electrons is much smaller than
the number of bound electrons.

\subsubsection{Ar single ionization}

\begin{figure}
\begin{center}
\includegraphics*[width=30pc]{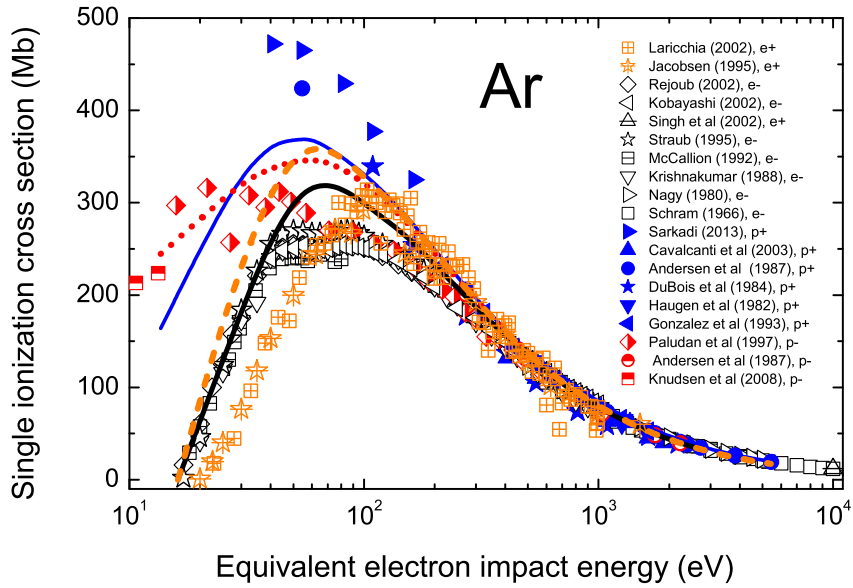}
\end{center}
\caption{\small{Single ionization of Ar by $|Z|=1$ projectiles as a
function of the impact energy, considering equal velocity for heavy
and light particles. Curves: CDW-EIS results for proton (blue thin
solid line), antiproton (red dotted line), electron (black thick
solid line) and positron (orange dashed line) impact. Symbols:
details in the inset; the references are: for protons p+
\cite{Cavalcanti03,haugen,Dubois84,Andersen87,gonzalez,sarkadi}, for
electrons e-
\cite{schram,nagy,krish,Rejoub,kobayashi,mccallion,straub,singh},
for positrons e+ \cite{Jacobsen95,Laricchia02}, and for antiprotons
p-  \cite{Andersen87,Paludan97,knudsenPRL08}.}\label{fig1Ar} }
\end{figure}

In figure \ref{fig1Ar} the single ionization of Ar by the four
$|Z|=1$ projectiles is displayed. The good description of the
experimental values by the CDW-EIS can be noted. The proton impact
measurements by Andersen {\textit et al}\cite{Andersen87} and by
Sarkadi {\textit et al} \cite{sarkadi} at impact velocities below
2.5 a.u. are clearly higher than the theoretical values. It may be
possible that these single ionization measurements by proton impact
include capture, enhancing the number of measured Ar$^+$. Obviously,
the antiproton measurements are pure ionization, and capture is not
possible.

In the high energy region, the convergence of the proton,
antiproton, positron and electron measurements and also of the
theoretical curves, is clearly above 300 eV. Again, our model
predicts that the ionization by positrons tends to antiproton and
proton values at lower energies than to the electron impact values.
In the case of Ar, this tendency is also found in the experimental
data by Laricchia and coworkers \cite{Laricchia02,Jacobsen95}.

\subsubsection{Ar double ionization}

\begin{figure}
\begin{center}
\includegraphics*[width=30pc]{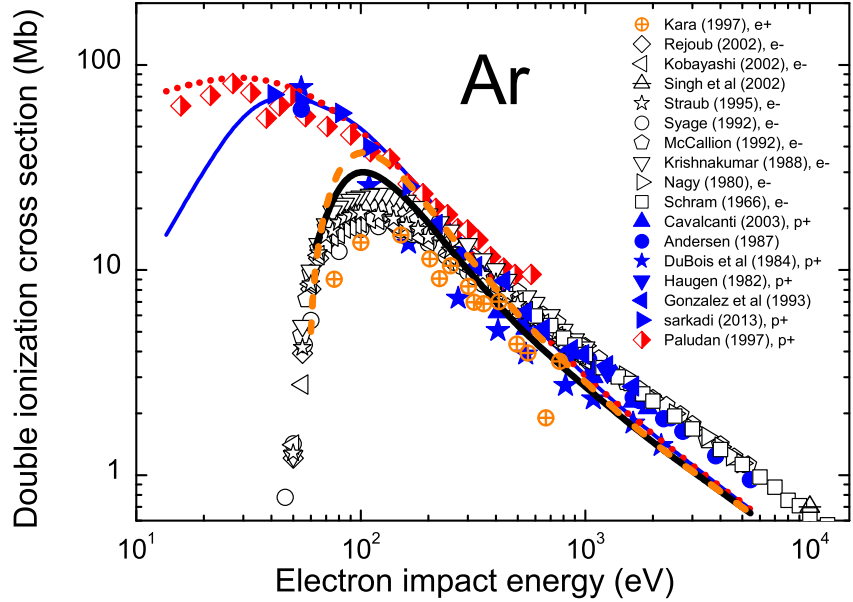}
\end{center}
\caption{\small{Double ionization of Ar by $|Z|=1$ projectiles as a
function of the impact energy, considering equal velocity for heavy
and light particles. Curves: CDW-EIS results for proton (blue thin
solid line), antiproton (red dotted line), electron (black thick
solid line) and positron (orange dashed line) impact. Symbols:
details in the inset; the references are: for protons p+
\cite{DuboisPRL84,Cavalcanti02,Cavalcanti03,haugen,Dubois84,DuboisM,Manson87,Andersen87,gonzalez},
for electrons e-
\cite{schram,nagy,syage,krish,Rejoub,kobayashi,mccallion,straub,singh},
for positrons e+ \cite{Kara97}, and for antiprotons p-
\cite{Paludan97}}.\label{fig2Ar} }
\end{figure}

In figure \ref{fig2Ar} the double ionization of Ar is shown. The
general description obtained with the CDW-EIS for the heavy and the
light projectiles is good. Though this model is not capable of
describing the low energy processes, the results are quite good in
this region too. The threshold for double ionization of Ar by
electron and positron impact is correctly described, as compared to
the data.

At high energies ($E>1$ keV) our curves are below most of the
electron-impact and proton-impact data. It is worth to mention that
all the Ar shells have been included in this calculation, even the
K-shell. So this unexplained undervalue of the theory may be related
to the empirical branching ratios employed.

\subsubsection{Ar triple ionization}

\begin{figure}
\begin{center}
\includegraphics*[width=30pc]{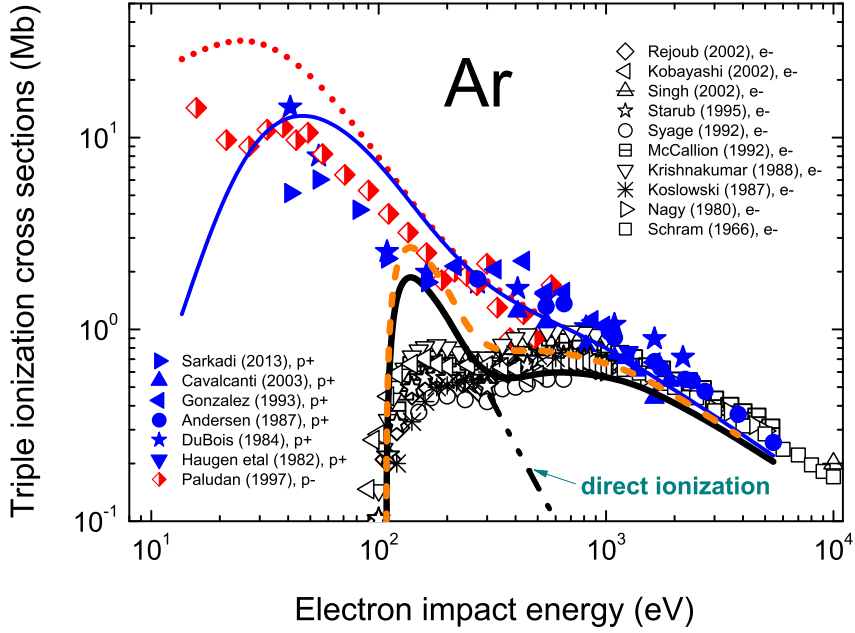}
\end{center}
\caption{\small{Triple ionization of Ar by $|Z|=1$ projectiles as a
function of the impact energy, considering equal velocity for heavy
and light particles. Curves: CDW-EIS results for proton (blue thin
solid line), antiproton (red dotted line), electron (black thick
solid line) and positron (orange dashed-line) impact; black
dashed-double-dotted line, direct triple ionization by electron
impact. Symbols: details in the inset; the references are: for
protons p+
\cite{Cavalcanti03,haugen,Dubois84,Andersen87,gonzalez,sarkadi}, for
electrons e-
\cite{schram,nagy,syage,krish,Rejoub,kobayashi,mccallion,straub,singh,koslowski},
and for antiprotons p- \cite{Paludan97}.}\label{fig3Ar} }
\end{figure}

In figure \ref{fig3Ar} the triple ionization of Ar is displayed. The
experimental data for heavy projectiles is rather well described,
though overestimated around the maximum. The theory predicts
antiproton values above the proton ones at the maximum of the cross
section. Experimentally, the data for protons match well with the
data for antiprotons.

In this figure we remark the importance of the PCI by including the
direct triple ionization values (dashed double-dotted line) and the
triple ionization including PCI (solid line) for electron impact
ionization. As can be noted, PCI is correctly included.  It is
interesting to note a double-shoulder shape in the triple ionization
by light projectiles. This is due to the passage from the energy
region of direct ionization (only valence-shell ionization) to the
region of  higher energies, where PCI is not negligible (inner shell
ionization). Note that this double shoulder can be observed also in
the electron impact measurements. The theory reproduces the shape
but overestimates below 250 eV. It would be interesting to have
measurements of triple ionization by positron impact to study this
effect too.

\begin{figure}
\begin{center}
\includegraphics*[width=30pc]{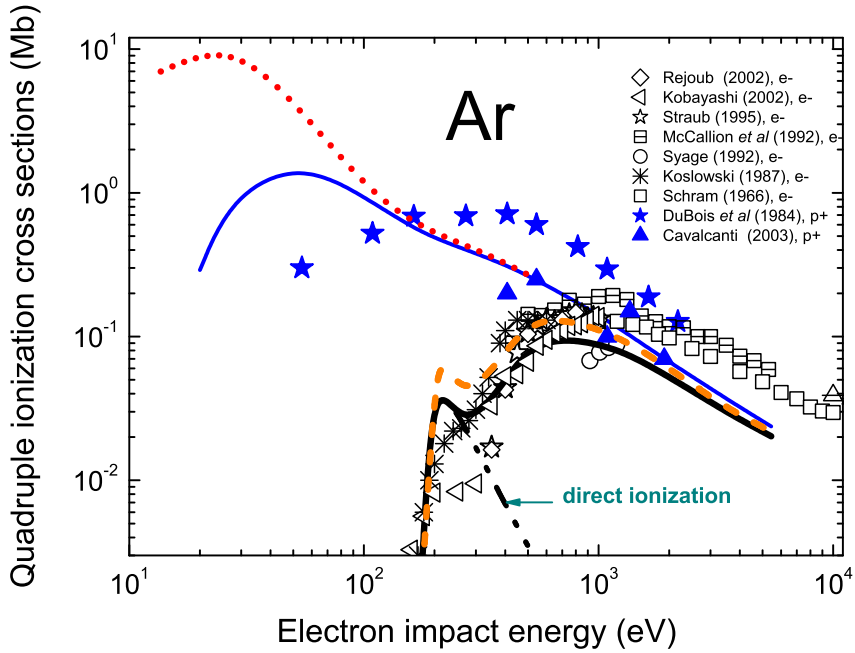}
\end{center}
\caption{\small{Quadruple ionization of Ar by $|Z|=1$ projectiles as
a function of the impact energy, considering equal velocity for
heavy and light particles. Curves: CDW-EIS results for proton (blue
thin solid line), antiproton (red dotted line), electron (black
thick solid line) and positron (orange dashed-line) impact. Symbols:
details in the inset; the references are: for protons p+
\cite{Cavalcanti03,Dubois84}, and for electrons e-
\cite{schram,syage,Rejoub,kobayashi,mccallion,straub,singh,koslowski}.}\label{fig4Ar}
}
\end{figure}

\subsubsection{Ar quadruple ionization}

In figure \ref{fig4Ar} the quadruple ionization cross sections of Ar
are displayed. Again we emphasize the importance of PCI by including
the direct quadruple ionization for the electron-impact case (dashed
double-dotted curve). Above 300 eV the PCI is crucial, instead the
direct ionization falls down, being negligible for impact energies
above 400 eV. The theory predicts a small shoulder around 200 eV
related to these two different ionization mechanisms. It can also be
noted that the threshold for electron-impact quadruple ionization is
nicely described.

Figure \ref{fig4Ar} shows that there are no antiparticle data for
quadruple ionization of Ar. This is related to the difficulties to
measure such low values, i. e. around or less than $10^{-18}$ cm$^2$
for proton impact, around or less than $10^{-19}$ cm$^2$ for
electron impact.

\begin{figure}
\begin{center}
\includegraphics*[width=30pc]{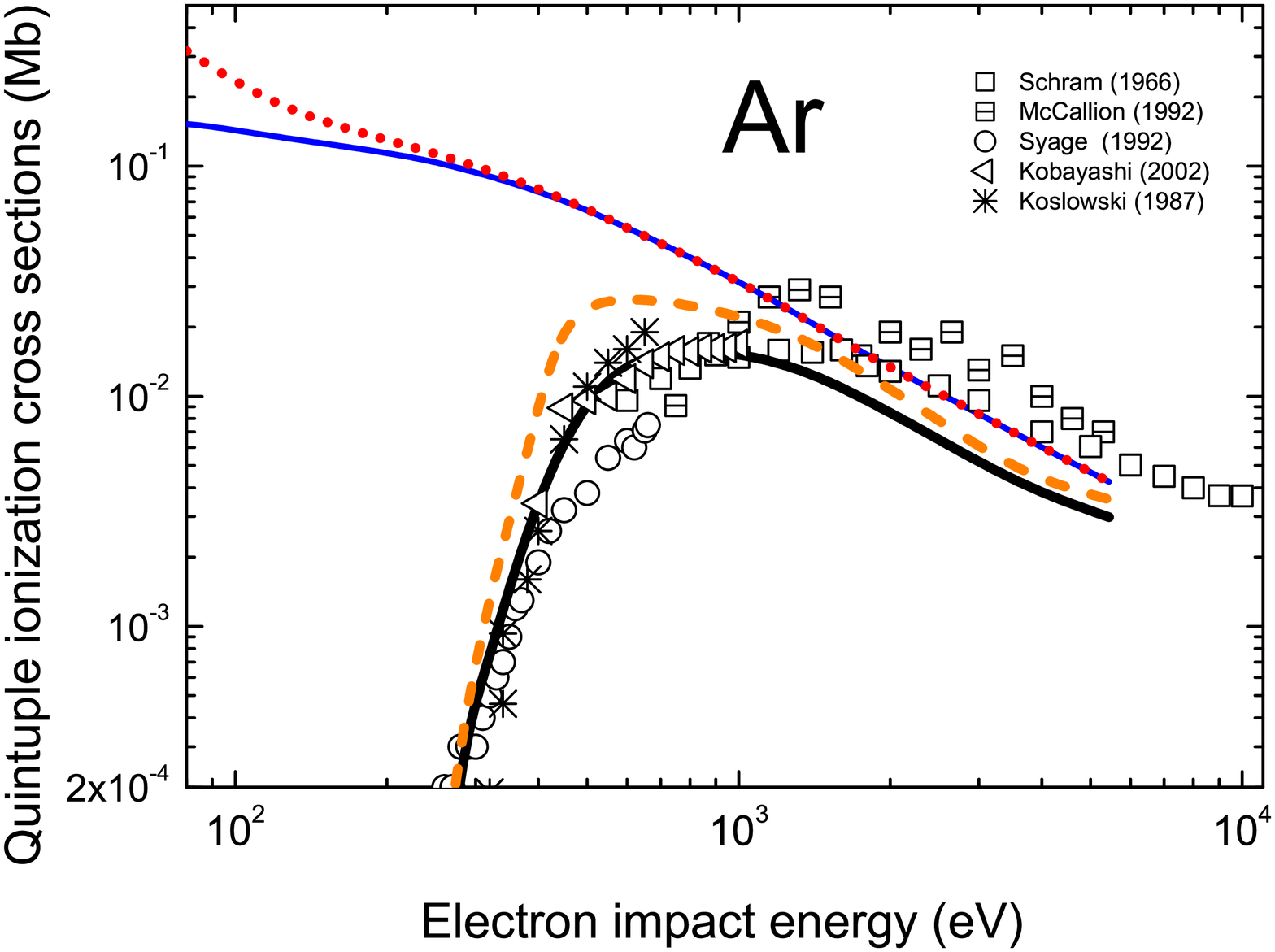}
\end{center}
\caption{\small{Quintuple ionization of Ar by $|Z|=1$ projectiles as
a function of the impact energy, considering equal velocity for
heavy and light particles. Curves: CDW-EIS results for proton (blue
thin solid line), antiproton (red dotted line), electron (black
thick solid line) and positron (orange dashed-line) impact. Symbols:
details in the inset; the references for electrons e- are
\cite{schram,syage,kobayashi,mccallion,singh,koslowski}.}\label{fig5Ar}
}
\end{figure}

\subsubsection{Ar quintuple ionization}

Finally in figure \ref{fig5Ar} the quintuple ionization of Ar is
displayed. The only data available in the literature is for
electron-impact quintuple ionization. Nevertheless we include the
CDW-EIS results for the four $|Z|=1$ projectiles. Hopefully, these
cross sections could be tested with future measurements by proton,
antiproton and positron impact.

The theoretical description is very good considering it is quintuple
ionization of Ar in an IPM approximation. The prediction of the
threshold at 230 eV is also fine. In this figure, the cross sections
by electron and positron impact are entirely due the PCI following
inner-shell ionization (1s, 2s, and 2p). And this is valid in the
whole energy range, even just above the threshold. The
direct-ionization contribution is so small that it is out of scale
in figure \ref{fig5Ar}, i.e. less than $2\ 10^{-22}$ cm$^2$.

\begin{figure}
\begin{center}
\includegraphics*[width=30pc]{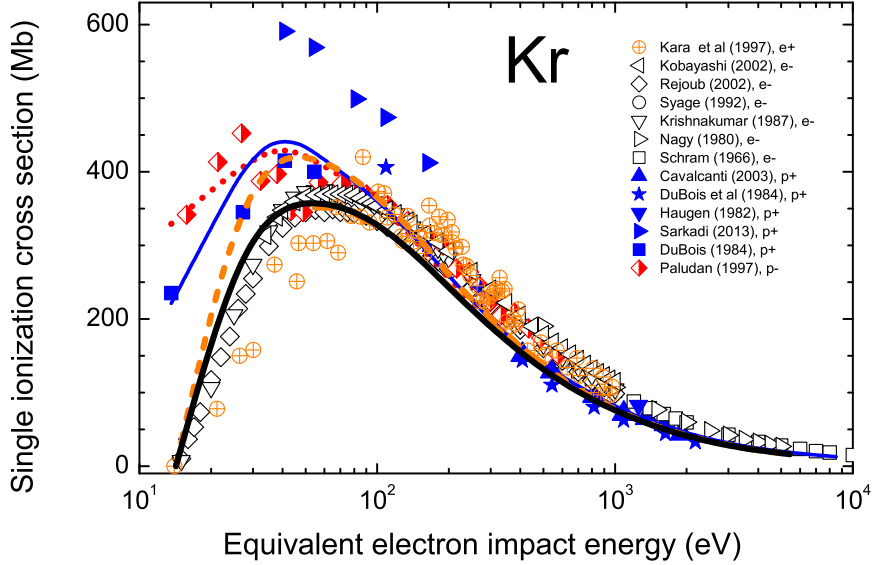}
\end{center}
\caption{\small{Single ionization of Kr by $|Z|=1$ projectiles as a
function of the impact energy, considering equal velocity for heavy
and light particles. Curves: CDW-EIS results for proton (blue thin
solid line), antiproton (red dotted line), electron (black thick
solid line) and positron (orange dashed-line) impact. Symbols:
details in the inset; the references are: for protons p+
\cite{DuboisPRL84,Cavalcanti03,haugen,Dubois84,sarkadi}, for
electrons e- \cite{schram,nagy,syage,krish,Rejoub,kobayashi}, for
positrons e+ \cite{Kara97}, and for antiprotons p-
\cite{Paludan97}.}\label{fig1Kr} }
\end{figure}

\subsection{The multiple ionization of Kr} \label{s4}

\begin{figure}
\begin{center}
\includegraphics*[width=30pc]{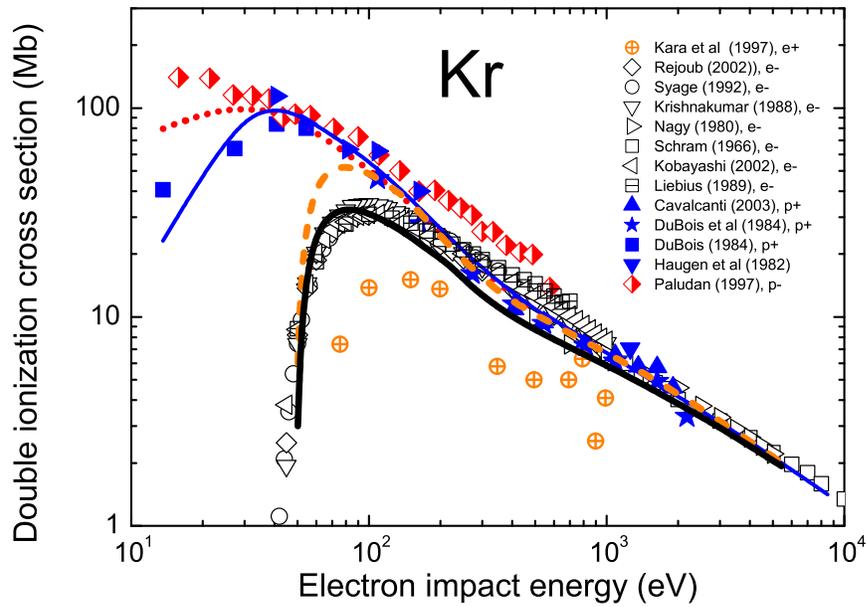}
\end{center}
\caption{\small{Double ionization of Kr by $|Z|=1$ projectiles as a
function of the impact energy, considering equal velocity for heavy
and light particles. Curves: CDW-EIS results for proton (blue thin
solid line), antiproton (red dotted line), electron (black thick
solid line) and positron (orange dashed-line) impact. Symbols:
details in the inset; the references are: for protons p+
\cite{DuboisPRL84,Cavalcanti02,Cavalcanti03,haugen,Dubois84,DuboisM,Manson87,Andersen87,gonzalez},
for electrons e-
\cite{schram,nagy,syage,krish,Rejoub,kobayashi,liebius}, for
positrons e+ \cite{Kara97}, and for antiprotons p-
\cite{Paludan97}.}\label{fig2Kr} }
\end{figure}

We present in this section a comparative study of Kr multiple
ionization by the four $|Z|=1$ projectiles, covering from single up
to quintuple ionization. In figures \ref{fig1Kr}-\ref{fig5Kr} we
displayed together the CDW-EIS results for proton, antiproton,
electron and positron \cite{AP,e-rare,positron}, and the available
experimental data.

\subsubsection{Kr single ionization}

In figure \ref{fig1Kr} we display the single ionization cross
sections of Kr. The theoretical-experimental agreement is very good.
Note that for electron impact the theoretical description agrees in
the maximum and in the energy threshold. Above 100 eV all the data
seem to converge. Similarly to the case of Ar, the data by Sarkadi
\textit{et al} \cite{sarkadi} for proton impact are quite above the
CDW-EIS predictions. But in this case we can also compare with the
measurements by DuBois and coworkers \cite{DuboisPRL84,DuboisM}.
DuBois experiments separate capture and pure ionization. The
differences between Sarkadi \textit{et al} \cite{sarkadi} and DuBois
data \cite{DuboisPRL84,DuboisM} below 60 eV are clear. This seems to
confirm the presence of capture in \cite{sarkadi}.

\subsubsection{Kr double ionization}

In figure \ref{fig2Kr} we can observe and analyze the double
ionization of Kr by particle and antiparticle impact. The good
description of Kr double ionization by antiproton, proton and
electron impact is remarkable. The theoretical and experimental
values show clearly that the cross sections for heavy projectiles
are quite similar for impact velocities above 1.7 a.u.

The theoretical results show that the double ionization cross
sections by the light projectiles are much lower than those of the
heavy projectiles at intermediate energies, with the threshold
around 40 eV. These values reproduce the electron impact data rather
well. However, the positron data by Kara \textit{et al}
\cite{Kara97} are below the theoretical predictions and quite below
the electron data.

At low impact energies the mass-effect dominates. The cross sections
by light projectiles are below the heavy projectiles ones, with the
characteristic sharp threshold. This contrasts with the
charge-effect at high energies. Above 100 eV positron values are
closer to proton values than to electron ones. The high energy
convergence of the different curves and data is evident at high
energies, as expected.

\subsubsection{Kr triple ionization}

\begin{figure}
\begin{center}
\includegraphics*[width=30pc]{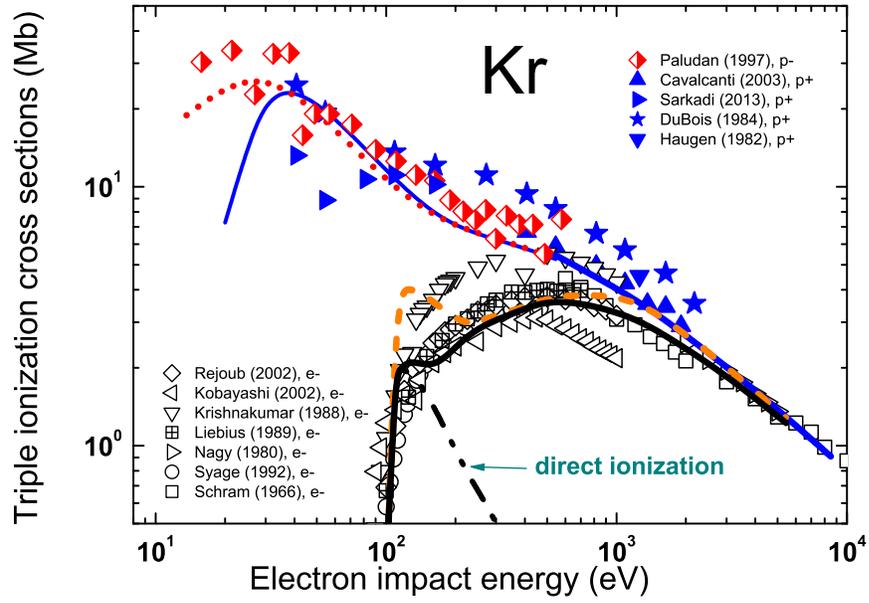}
\end{center}
\caption{\small{Triple ionization of Kr by $|Z|=1$ projectiles as a
function of the impact energy, considering equal velocity for heavy
and light particles. Curves: CDW-EIS results for proton (blue thin
solid line), antiproton (red dotted line), electron (black thick
solid line) and positron (orange dashed-line) impact. Symbols:
details in the inset; the references are: for protons p+
\cite{Cavalcanti03,haugen,Dubois84,sarkadi}, for electrons e-
\cite{schram,nagy,syage,krish,Rejoub,kobayashi,liebius}, and for
antiprotons p- \cite{Paludan97}.}\label{fig3Kr} }
\end{figure}

The theoretical results presented in figure \ref{fig3Kr} nicely
describe the measurements of proton, antiproton and electron impact.
No measurements for triple ionization of Kr by positrons have been
reported yet. For light projectiles, PCI is the main contribution
almost in the whole energy range. This is remarked in figure
\ref{fig3Kr} with a separate curve that shows the direct triple
ionization without PCI. Again, the small hump near the threshold is
associated with the appearance of the PCI contribution. It is
present in the theoretical curve for electron impact, and may be
noted in the certain data. The bigger hump for positron impact could
be an artifact of the theoretical calculations. However, no
theoretical experimental comparison is possible for positron impact.

As already mentioned, the mass effect is more important than the
charge effect at low and intermediate energies, while equal charge
determines the energy for which the curves converge. Theoretically,
the positron cross sections tend to proton ones for $E> 1$ keV.
Instead electron impact cross sections converge to the rest above 2
keV. This may indicate that when the ionization of inner shells
dominates, the mass effect prevails over the charge effect.

\subsubsection{Kr quadruple ionization}

\begin{figure}
\begin{center}
\includegraphics*[width=30pc]{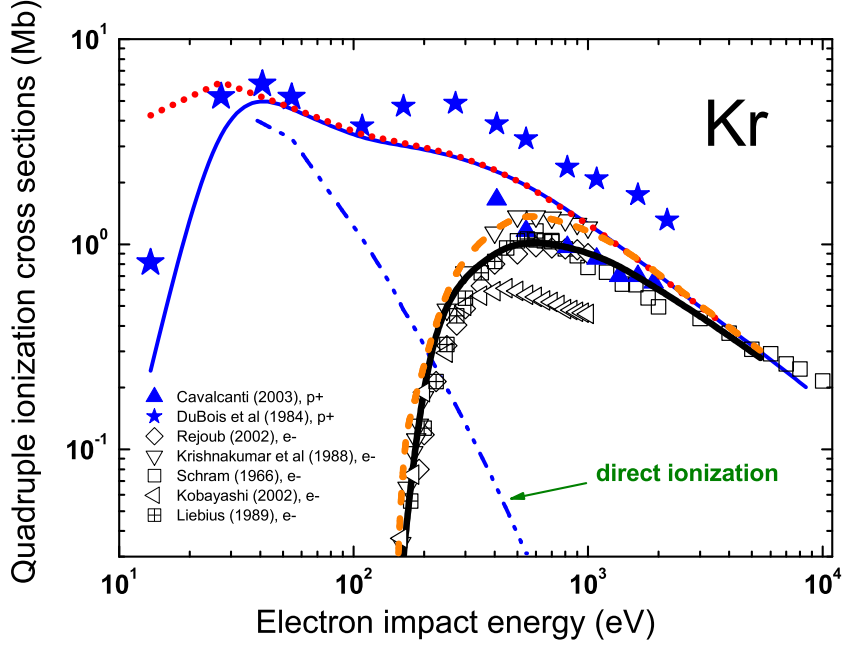}
\end{center}
\caption{\small{Quadruple ionization of Kr by $|Z|=1$ projectiles as
a function of the impact energy, considering equal velocity for
heavy and light particles. Curves: CDW-EIS results for proton (blue
thin solid line), antiproton (red dotted line), electron (black
thick solid line) and positron (orange dashed-line) impact. Symbols:
details in the inset; the references are: for protons p+
\cite{Cavalcanti03,Dubois84}, and for electrons e-
\cite{schram,krish,Rejoub,kobayashi,liebius}}.\label{fig4Kr} }
\end{figure}

Figure \ref{fig4Kr} shows the quadruple ionization of Kr. The
theoretical-experimental agreement is noticeable. The comparison is
very interesting. The direct ionization is a minor contribution for
light particles even near the energy threshold. In the case of heavy
particles, PCI is important at rather low energies. This is remarked
in figure \ref{fig4Kr} by including the theoretical direct quadruple
ionization cross section by proton impact. Below 700 eV the
theoretical results for positron impact are close to the electron
ones, but no comparison with the experiments is possible for
positron impact. The lack of antiparticle cross sections is evident
in this figure. We expect that the present study may be of interest
for future measurements.

\subsubsection{Kr quintuple ionization}

\begin{figure}
\begin{center}
\includegraphics*[width=30pc]{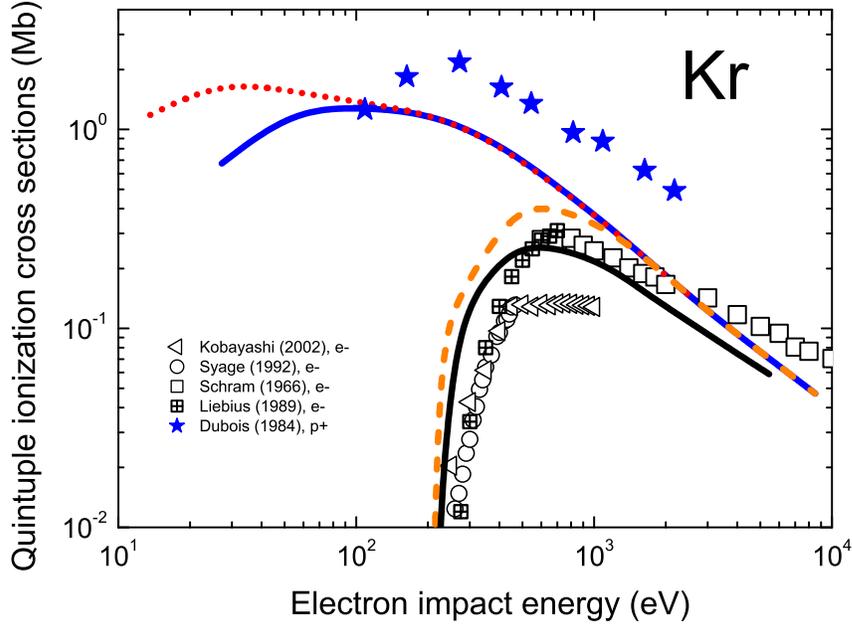}
\end{center}
\caption{\small{Quintuple ionization of Kr by $|Z|=1$ projectiles as
a function of the impact energy, considering equal velocity for
heavy and light particles. Curves: CDW-EIS results for proton (blue
thin solid line), antiproton (red dotted line), electron (black
thick solid line) and positron (orange dashed-line) impact. Symbols:
details in the inset; the references are: for protons p+
\cite{Dubois84}, and for electrons e-
\cite{schram,syage,kobayashi,liebius}.}\label{fig5Kr} }
\end{figure}

Finally, the quintuple ionization is displayed in figure
\ref{fig5Kr}. The behavior is similar to the one observed for
quadruple ionization. Note that for proton impact there is only one
set of experimental data, by DuBois \cite{DuboisPRL84}. The case is
different for electron impact measurements, which are by far the
most abundant of the four projectiles studied here. However, we
included the theoretical predictions for the four $|Z|=1$
projectiles in view of future measurements.

For electron impact, PCI dominates the quintuple ionization in the
whole energy range. The direct quintuple ionization is negligible
even for energies near the threshold. It is worth to mention that to
describe the quintuple ionization of Kr, even the ionization of the
L-shell must be considered \cite{sigaud13}. Moreover, the good
performance of the CDW-EIS for the Kr$^{+5}$ final charge state is
due to the correct description of the ionization probabilities of
the very deep shells, both by heavy and light projectiles.

\subsection{The multiple ionization of Xe}\label{s5}

\begin{figure}
\begin{center}
\includegraphics*[width=35pc]{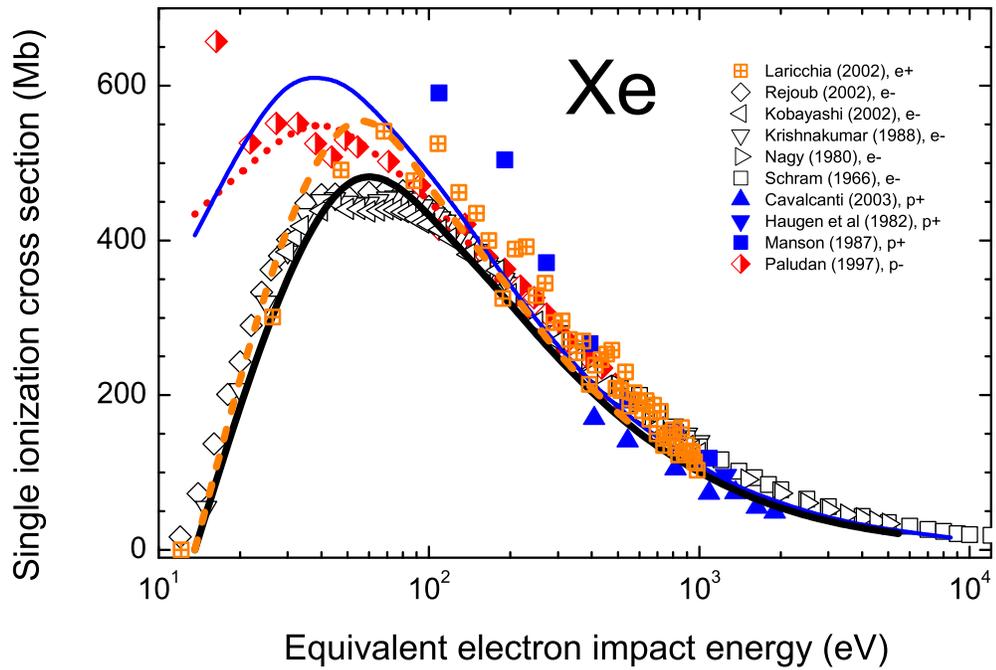}
\end{center}
\caption{\small{Single ionization of Xe by $|Z|=1$ projectiles as a
function of the impact energy, considering equal velocity for heavy
and light particles. Curves: CDW-EIS results for proton (blue thin
solid line), antiproton (red dotted line), electron (black thick
solid line) and positron (orange dashed-line) impact. Symbols:
details in the inset; the references are: for protons p+
\cite{Cavalcanti03,haugen,Manson87}, for electrons e-
\cite{schram,nagy,syage,krish,Rejoub,kobayashi}, for positrons e+
\cite{Laricchia02}, and for antiprotons p-
\cite{Paludan97}.}\label{fig1Xe} }
\end{figure}

Finally, Xenon multiple ionization by $|Z=1|$ projectiles is
displayed in figures \ref{fig1Xe}-\ref{fig5Xe}, from single to
quintuple ionization. The results for multiple ionization of Xe in
these figures show the scope of the model that combines CDW-EIS, IPM
for multiple ionization, mass effects in the description of light
projectiles and the PCI through the empirical branching ratios. The
theoretical results for Xe show a very good description of the
experimental measurements for the different $|Z|=1$ projectiles,
meaning that the charge effects (positive vs. negative charges,
repulsive or attractive potential) and the mass effects (difference
in the impact energy, threshold, non-linear trajectories) are
correctly included. On the other hand, it is reasonable that the IPM
approximation (no electron-electron correlation, no changes in the
target potential due to the loss of one or more electrons) works
better for Xe with 54 electrons than for Ne, with only 10 electrons.
The Xe potential is less affected by the loss of a few electrons,
and the approximation of independent events without changes is more
realistic.

\subsubsection{Xe single ionization}

In Figure \ref{fig1Xe} the ionization of just one electron is
displayed in linear scale in order to show in detail the
measurements and the theoretical cross sections around the maximum.
The CDW-EIS predicts that proton and antiproton maximum cross
sections are around 40 eV (in electron impact energy, it is $v \sim
1.7$ a.u.), while positron and electron curves are shifted to 55-60
eV. It is very interesting that the positron and antiproton highest
cross sections are almost equal. The electron impact values are
below the positron data around the maximum. For impact energies
40-80 eV, the single ionization by positron-impact is more effective
(higher probabilities) than by electron-impact.

The very good agreement with the experimental values reinforce these
conclusions. The measurements for the different projectiles converge
at high energies, and show the peculiar behavior for each projectile
at lower energies. The description of the antiproton measurements by
Paludan \textit {et al} \cite{Paludan97} is quite good in the whole
energy range. The same for electron and positron impact. Clearly the
positron measurements are above the electron ones, as predicted
theoretically, with values very similar to those of antiproton. The
experimental data for proton impact by Manson {\textit et al}
\cite{Manson87} are well described at high energies, but the three
measurements for impact velocities $v<3$ a.u. are clearly above the
theoretical curve. The comparison of them with the single ionization
by antiproton, which is actually pure ionization, suggests that
electron capture may be present in this proton impact data. At high
energies, the measurements of single ionization of Xe by proton
impact by Cavalcanti \textit{et al} \cite{Cavalcanti03} are below
the rest. It would be useful to compare measurements for Xe$^+$
production by proton impact at $1 \leq v \leq 4$  with the similar
antiproton values.

\subsubsection{Xe double ionization}

\begin{figure}
\begin{center}
\includegraphics*[width=30pc]{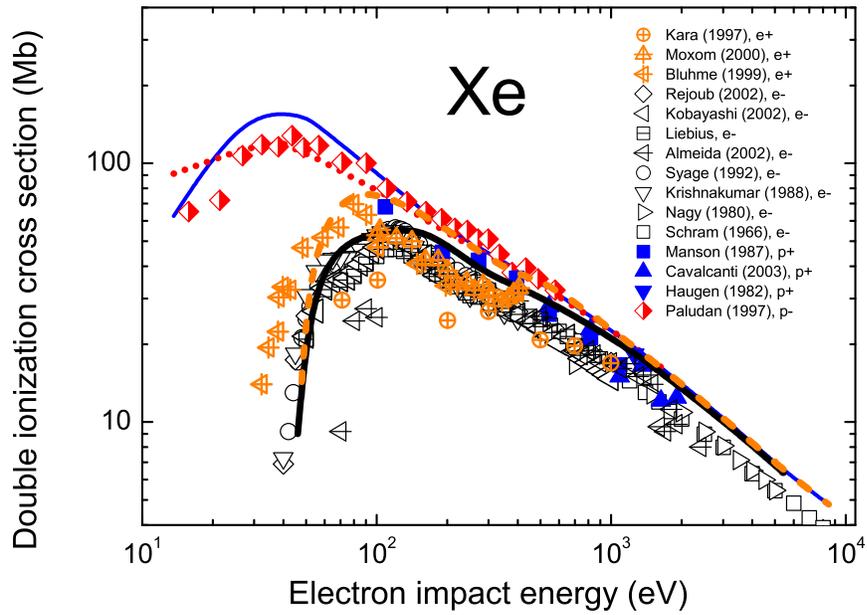}
\end{center}
\caption{\small{Double ionization of Xe by $|Z|=1$ projectiles as a
function of the impact energy, considering equal velocity for heavy
and light particles. Curves: CDW-EIS results for proton (blue thin
solid line), antiproton (red dotted line), electron (black thick
solid line) and positron (orange dashed-line) impact. Symbols:
details in the inset; the references are: for protons p+
\cite{Cavalcanti03,haugen,Manson87}, for electrons e-
\cite{schram,nagy,syage,krish,Rejoub,kobayashi,almeida,liebius}, for
positrons e+ \cite{Kara97,Moxom00,bluhme99}, and for antiprotons p-
\cite{Paludan97}.}\label{fig2Xe} }
\end{figure}

The agreement between the theoretical and the experimental values in
figure \ref{fig2Xe} is amazing. These cross sections are not so
small, i.e. around $10^{-15} \ - \ 10^{-16}$ cm$^2$. The
experimental values for double ionization of Xe by antiprotons are
very interesting because they cover the energy range around the
maximum of the cross section. Instead, there is a lack of proton
impact data in this energy region (velocity $v < 3$ a.u.). For
positron impact, there are three groups of independent data with
rather good agreement with the CDW-EIS results. It may be noted that
the low energy measurements by Bluhme \textit{et al}\cite{bluhme99}
suggest a different energy threshold for electron and positron
impact. This is not shown by the theory \cite{hci,positron} included
here. However, it is worth to remember here that the CDW-EIS is not
expected to be valid near the energy threshold. So we cannot state a
conclusion abut this positron-electron difference.

Above 1 keV all the double ionization cross sections converge. These
values are highly influenced by the single ionization of a
subvalence electron followed by decay and emission of a second
electron. This PCI is important for energies above 200 eV and
represents $80\%$ of the double ionization cross section at 1 keV.

\begin{figure}
\begin{center}
\includegraphics*[width=30pc]{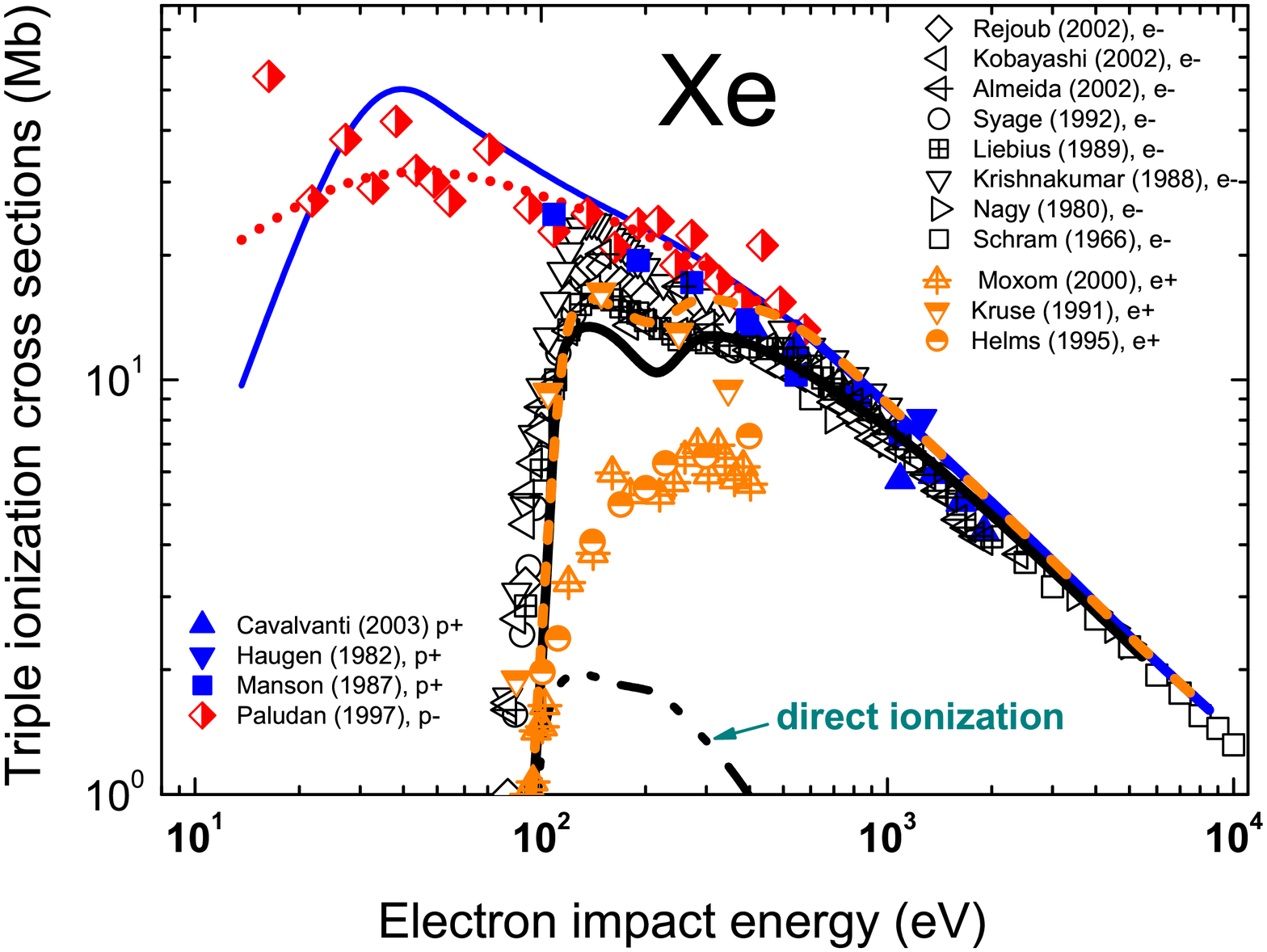}
\end{center}
\caption{\small{Triple ionization of Xe by $|Z|=1$ projectiles as a
function of the impact energy, considering equal velocity for heavy
and light particles. Curves: CDW-EIS results for proton (blue thin
solid line), antiproton (red dotted line), electron (black thick
solid line) and positron (orange dashed-line) impact. Symbols:
details in the inset; the references are: for protons p+
\cite{Cavalcanti03,haugen,Manson87}, for electrons e-
\cite{schram,nagy,syage,krish,Rejoub,kobayashi,almeida,liebius}, for
positrons e+ \cite{Moxom00,helms,kruse}, and for antiprotons p-
\cite{Paludan97}.}\label{fig3Xe} }
\end{figure}

\subsubsection{Xe triple ionization}

The triple ionization of Xe, theory and compilation of available
data, is presented in figure \ref{fig3Xe}.  The comparison shows
again the validity of the description. The antiproton, proton and
electron impact data are in reasonable agreement with the
theoretical curves. For triple ionization, the threshold for light
projectile ionization is very sharp and pushes down the cross
sections. Theoretically, the cross sections are similar for electron
and positron impact, with the two humps and a small minimum at 200
eV.  At high energies, the positron curve converges to proton one at
400 eV, while the electron curve does above 1 keV. This high energy
triple ionization is mostly PCI, as indicated in the figure. The
curve obtained just considering the direct triple ionization
(without PCI) is far from describing the data.

It can be noted in figure \ref{fig3Xe} that there are differences
among the three sets of positron data. The measurements by Kruse
{\textit et al} \cite{kruse} agree with the expected values around
the maximum. However, the more recent measurements by Moxom
\textit{et al} \cite{Moxom00} and Helms \textit {et al} \cite{helms}
are below the curves and the electron-impact data. Again we can draw
attention to the lack of proton impact measurements at intermediate
and low energies.

\subsubsection{Xe quadruple ionization}

\begin{figure}
\begin{center}
\includegraphics*[width=30pc]{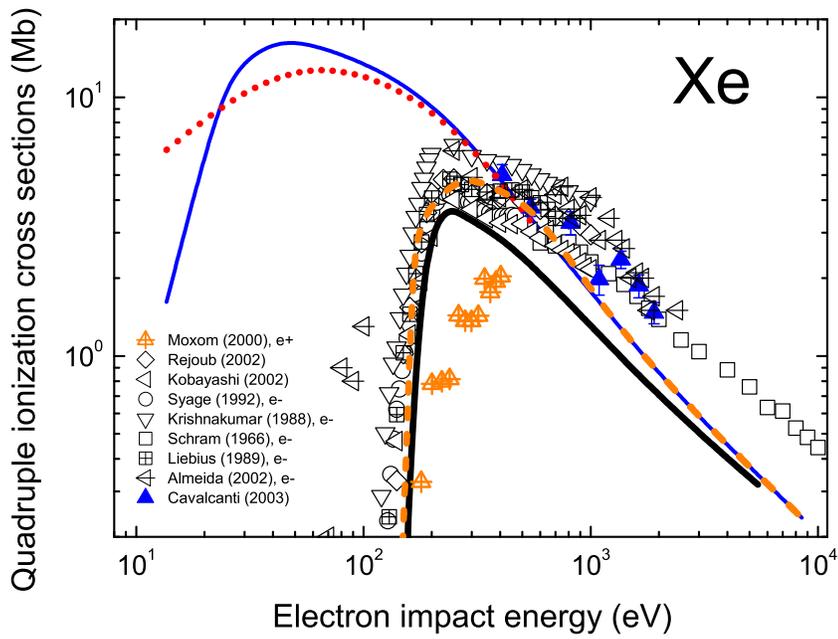}
\end{center}
\caption{\small{Quadruple ionization of Xe by $|Z|=1$ projectiles as
a function of the impact energy, considering equal velocity for
heavy and light particles. Curves: CDW-EIS results for proton (blue
thin solid line), antiproton (red dotted line), electron (black
thick solid line) and positron (orange dashed-line) impact. Symbols:
details in the inset; the references are: for protons p+
\cite{Cavalcanti03}, for electrons e-
\cite{schram,syage,krish,Rejoub,kobayashi,almeida,liebius}, and for
positrons e- \cite{Moxom00}}.\label{fig4Xe} }
\end{figure}

In the figure \ref{fig4Xe} the quadruple ionization of Xe is
displayed, including the experimental data available and the CDW-EIS
results for the four $|Z|=1$ projectiles. The abundance of electron
impact data can be noted, while there is only one set for positron
impact \cite{Moxom00}, and only one set for proton impact
\cite{Cavalcanti03}. No antiproton values are available in the
literature, and proton data is available only in the energy region
where they converge to electron-impact ones. The lack of
measurements for the heavy projectiles around the maximum of the
cross section does not allow to evaluate the theoretical predictions
for proton and antiprotons. On the other hand, the positron
measurements by Moxom {\textit et al} \cite{Moxom00} are a factor 2
below the theoretical values.

The agreement with the electron impact data below 300 eV is fine
considering they are quadruple-ionization cross sections. On the
contrary, the theoretical high-energy values are too small.
Concerning this undervalue, the CDW-EIS results in \cite{positron}
include very deep shells, so the branching ratios of PCI may be the
reason. However, it is fair to remark that at 5 keV the direct
quadruple ionization cross sections are 4 orders of magnitude below
the experimental values. Thus, the theoretical values including PCI,
even a factor two below the data at 5 keV, are a rather good
description of the electron impact values.

\subsubsection{Xe quintuple ionization}

\begin{figure}
\begin{center}
\includegraphics*[width=30pc]{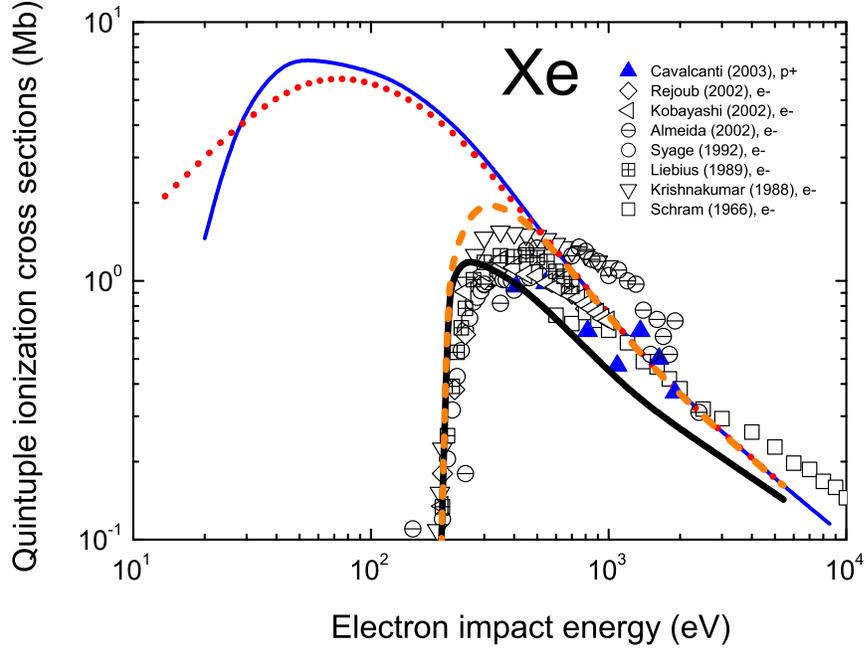}
\end{center}
\caption{\small{Quintuple ionization of Xe by $|Z|=1$ projectiles as
a function of the impact energy, considering equal velocity for
heavy and light particles. Curves: CDW-EIS results for proton (blue
thin solid line), antiproton (red dotted line), electron (black
thick solid line) and positron (orange dashed-line) impact. Symbols:
details in the inset; the references are: for protons p+
\cite{Cavalcanti03}, and for electrons e-
\cite{schram,syage,krish,Rejoub,kobayashi,almeida,liebius}.}\label{fig5Xe}
}
\end{figure}

Finally, for quintuple ionization of Xe in figure \ref{fig5Xe}, only
electron and proton impact data are available in the literature. For
electron impact, the theoretical curve is in good agreement with the
experimental measurements, considering the difficulties of such low
values. The high energy proton measurements by Cavalcanti {\textit
et al} \cite{Cavalcanti03} also agree with the electron data, except
with those by Almeida {\textit et al} \cite{almeida} that are larger
than the rest. The theoretical curves for proton and antiproton
impact are quite similar around the maximum. But there is no
experimental data to confirm this.

There are some characteristics to remark about the CDW-EIS results
for quintuple ionization displayed in figure \ref{fig5Xe}: the first
one is that at low impact energies, the energy threshold for
electron and positron impact is nicely described. The inclusion of
the mean energy transferred in the threshold for multiple ionization
proved to be important \cite{hci}. For example, while the outer 5s
and 5p electrons of Xe have binding energies around 13 eV, the
threshold for quintuple ionization is 178 eV \cite{positron}, much
greater than $5 \times 13$ eV. The second point to emphasize is that
the Auger-like processes leading to quintuple ionization are by far
the main contribution for light projectiles in the whole energy
range, even near the threshold. The single ionization of the Xe 4s,
3d, and 3p electrons are the most important contributions to final
quintuple ionization. This is clear in view of the experimental
branching ratios for Xe in table 4 of \cite{e-rare}. At high
energies, again the convergence of positron values to proton and
antiproton takes place around 400 eV, while electron impact values
do so only above 6 keV.

\newpage
\section{Concluding remarks and future prospects}\label{s6}

In this chapter the particle and antiparticle multiple ionization of
Ne, Ar, Kr and Xe was analyzed considering heavy and light
projectiles. This was done based on the CDW-EIS results and a vast
compilation of the experimental data available in the literature for
the sixteen collisional systems, four projectiles and four targets.

The theoretical formalism combines the independent particle model,
the CDW-EIS ionization probabilities for heavy projectiles, the
changes in this approximation to consider light-particle collisions
(difference in the impact energy, threshold, non-linear
trajectories), and the inclusion of the post-collisional
contribution to the final charge state due to Auger-like processes.

The multiple ionization of Kr and Xe shows the scope of the model:
good description for the different $|Z|=1$ projectiles, meaning that
the charge  and mass effects  are correctly included. The
improvement of the theoretical results with the target nuclear
charge (or with the number of bound electrons) is evident. The
employment of the independent particle model to obtain multiple
ionization probabilities does not consider changes in the target
potential due to the loss of one or more electrons. It is reasonable
that this approximation works better for atoms such as Kr or Xe than
for Ne.

The correction of the CDW-EIS approximation to describe light
particles shows the tendency of the experimental data. On the other
hand, the energy thresholds calculated separately and imposed within
the multinomial expansion, proved to describe rather well the
experimental values for electron and positron impact. Although the
threshold itself is rather well described, the CDW-EIS approximation
is not expected to be valid for low energies.

The highly-charged ion production is dominated by the
post-collisional electron emission that follows the inner-shell
ionization (rearrangement and/or Auger processes). The CDW-EIS
results are good even for quadruple and quintuple ionization cross
sections. These calculations are sensitive to the good description
of the deep shell ionization. They also rely on the empirical
branching ratios of post-collisional ionization. We concluded that
in multiple ionization by light particles, the post-collisional
contribution is the main ionization channel in the whole energy
range, even close to the threshold.

Future prospects can be separated in two lines. Within the
theoretical work, the case of Ne is one of the limitations of the
model. Not only related to the independent electron approximation,
but also to the inclusion of the shake-off. Future progress in this
regard is expected. On the other hand, the CDW-EIS has already been
tested for sextuple ionization of Kr and Xe by electron impact with
very good agreement with the experimental data. The extension to
higher ionization orders (final charge state $+q \geq 7$) requires
high computational effort but is possible.

Within the experimental research, the extensive compilation of data
and the comparison of electron, positron, proton and antiproton
values have been very enlightening. Unfortunately, the thorough
particle-antiparticle comparison could be made up to double
ionization. This is an interesting point to consider for future
research. There are no positron impact experiments for higher levels
of multiple ionization of rare gases. The exception is Xe. The
antiproton data available goes up to triple ionization. It is
interesting to note that, though the requirements and difficulties
of the antiproton experiments (high energy physics facility), there
are more measurements for ionization of Xe by antiprotons than by
protons.

Present review also discussed the normalization of antiparticle
relative measurements to the high energy electron ionization cross
sections. We found that at high energies, the ionization cross
sections by antiparticles impact converge to proton impact values at
lower energies rather than to electron impact ones. Recent
techniques for electron ionization measurements introduce very low
relative errors, and make them quite interesting and reliable for
the normalization purposes. The point to consider is that from which
impact energy positron and electron, or antiproton and electron
ionization cross sections, are equal.

\section*{Acknowledgments}

This work was partially supported by the following institutions in
Argentina: Consejo Nacional de Investigaciones Cient\'{\i}ficas y
T\'{e}cnicas, Agencia Nacional de Promoci\'{o}n Cient\'{\i}fica y
Tecnol\'{o}gica, and Universidad de Buenos Aires. I thank Jorge
Miraglia for useful discussions on the theoretical results.


\begin{thebibliography}{999}
\scriptsize{

\bibitem{McGuire88} McGuire J H 1986 \textit{Positron (Electron)-Gas
Scattering}, eds. Kauppila W E, Stein T S and Wadehra J M
(Singapore, World Scientific) 222-231

\bibitem{Schultz91} Schultz D R, Olson R E and Reinhold C 0 1991 \textit{J.
Phys. B: At. Mol. Opt. Phys.} \textbf{24} 521-558

\bibitem{knudsen} Knudsen H and Reading J F 1992, {\it Phys. Report} {\bf 212}
107-222

\bibitem{Knudsen92} Knudsen H and Reading J F 1992 \textit{Phys. Rep.}
\textbf{212} 107-222

\bibitem{P-L} Paludan K, Laricchia G, Ashley P, Kara V, Moxom J,
Bluhme H, Knudsen H, Mikkelsen U, M{\"o}ller S P, Uggerh{\"o}j U,
and Morenzoni E 1997, \textit{J. Phys. B: At. Mol. Opt. Phys.}
\textbf{30} L581-L587


\bibitem{DuboisPRL84} DuBois R D 1984, {\it Phys. Rev. Lett} {\bf 52},
2348-2351

\bibitem{Laricchia13} G Laricchia, Cooke D A, K{\"o}ver A, Brawley S
J 2013 \textit{Experimental Aspects of Ionization Studies by
Positron and Positronium Impact} (Cambridge University Press)
\textbf{56} 116-136



\bibitem{Cavalcanti02} Cavalcanti E G, Sigaud G M, Montenegro E C, Sant
'Anna M M, and Schmidt-Bocking H 2002 \textit{J. Phys. B: At. Mol.
Opt. Phys. }\textbf{35} 3937-3944

\bibitem{Cavalcanti03} Cavalcanti E G, Sigaud G M, Montenegro E C,
and Schmidt-Bocking H 2003 \textit{J. Phys. B: At. Mol. Opt. Phys.
}\textbf{36} 3087-3096

\bibitem{Spranger04} Spranger T and Kirchner T 2004 \textit{J. Phys. B: At.
Mol. Opt. Phys. }\textbf{37} 4159

\bibitem{Galassi07} Galassi M E, Rivarola R D, and Fainstein P D 2007,
{\it Phys. Rev. A} {\bf 75}, 052708

\bibitem{MM1} Montanari C C, Montenegro E C and Miraglia J E 2010, \textit{%
J. Phys. B: At. Mol. Opt. Phys.} \textbf{43}, 165201

\bibitem{e-rare} Montanari C C and Miraglia J E 2014, \textit{J. Phys. B:
At. Mol. Opt. Phys.} \textbf{47}, 105203.


\bibitem{haugen} Haugen H K, Andersen L H, Hvelplund P and Knudsen H 1982,
{\it Phys. Rev} A {\bf 26} 1962-1974

\bibitem{Dubois84} DuBois R D, Toburen L H and Rudd M E
1984 {\it  Phys. Rev. A }{\bf 29} 70-76

\bibitem{Andersen87} Andersen L H,
Hvelplund P, Knudsen H, M\"{o}ller S P, S\"{o}rensen A H, Elsener K,
Rensfelt K G and Uggerh\"{o}j 1987 \textit{Phys. Rev. A} \textbf{36}
3612-3629

\bibitem{DuboisM} DuBois R D and Manson S T 1987 {\it Phys. Rev.} A
{\bf 35} 2007-2025

\bibitem{Manson87} Manson S T and DuBois R D 1987 {\it J. Physique}
\textbf{48} C9 263-266 (online by EDP Sciences
http://dx.doi.org/10.1051/jphyscol:1987945


\bibitem{gonzalez} Gonzalez A D and Horsdal Pedersen E 1993, {\it Phys. Rev. A} %
{\bf 48} 3689

\bibitem{sarkadi} Sarkadi L, Herczku P, Kovacs S T S, and Kover A
2013, \textit{ Phys. Rev} A {\bf 87} 062705



\bibitem{schram} Schram B L, Boerboom A J H and Kistermaker J 1966 \textit{%
Physica} \textbf{32} 185-196; Schram B L 1966 \textit{Physica}
\textbf{32} 197-209; Schram B L, de Heer F J, Van der Wiel M J and
Kistermaker J 1965 \textit{\ Physica} \textbf{31} 94; Adamczyk B,
Boerboom A J H, Schram B L and Kistermaker J 1966 \textit{J. Chem.
Phys.} \textbf{44} 4640-4642

\bibitem{nagy} Nagy P, Skutlartz A and Schmidt V 1980, \textit{J. Phys. B:
At. Mol. Opt. Phys.} \textbf{13} 1249-1267
\bibitem{syage} Syage J A 1992, \textit{\ Phys. Rev.} A \textbf{46} 5666

\bibitem{krish} Krishnakumar E and Srivastava S K 1988, \textit{\ J. Phys.
B: At. Mol. Opt. Phys.} \textbf{21} 1055-1082


\bibitem{Rejoub} Rejoub R, Lindsay B G and Stebbing R F 2002, \textit{\
Phys. Rev.} A \textbf{65} 042713

\bibitem{kobayashi} Kobayashi A, Fujiki G, Okaji A and Masuoka T 2002
\textit{\ J. Phys. B: At. Mol. Opt. Phys. } \textbf{35} 2087-2103

\bibitem{mccallion} McCallion P, Shah M B and Gilbody H B 1992, \textit{\ J.
Phys. B: At. Mol. Opt. Phys.} \textbf{25} 1061-1071

\bibitem{straub} Straub H C, Renault P, Lindsay B G, Smith K A and Stebbings
R F 1995, \textit{Phys. Rev.} A \textbf{52} 1115-1124

\bibitem{almeida} Almeida D P, Fontes A C and Godinho C F L 1995, \textit{J.
Phys. B At. Mol. Opt. Phys.} \textbf{28} 3335-3345

\bibitem{liebius} Liebius H, Binder J, Koslowwski H R, Wiesemann K and Huber
B A 1989, \textit{J. Phys. B: At. Mol. Opt. Phys.} \textbf{22} 83-97

\bibitem{singh} Singh R K, Hippler R and Shanker R 2002, {\it J. Phys. B: At. Mol. Opt. Phys. }%
{\bf 35} 3243-3256.

\bibitem{koslowski} Koslowski H R, Binder J, Huber B A and Wiesemann
K 1987, \textit{J. Phys. B: At. Mol. Opt. Phys.} \textbf{20}, 5903



\bibitem{Andersen86} Andersen L H,
Hvelplund P, Knudsen H, M\"{o}ller S P, Elsener K, Rensfelt K G and
Uggerh\"{o}j 1986, \textit{Phys. Rev. Lett} \textbf{57} 2147-2150

\bibitem{Andersen89} Andersen L H, Hvelplund P, Knudsen H, M\"{o}ller S
P, Pedersen J O P, Tang-Pedersen S, Uggerh\"{o}j E, Elsener K and
Morenzoni E 1989, \textit{Phys. Rev. A} \textbf{40} 7366

\bibitem{Paludan97} Paludan K, Bluhme H, Knudsen H, Mikkelsen U, M\"{o}ller S
P, Uggerh\"{o}j E and Morenzoni E 1997, {\it  J. Phys. B: At. Mol.
Opt. Phys. }{\bf 30} 3951

\bibitem{knudsenPRL08} Knudsen H  {\textit et al} 2008, {\it Phys. Rev. Lett} {\bf 101} 043201
107-222

\bibitem{knudsenNIMB09} Knudsen H  {\textit et al} 2009, {\it Nucl. Instrum and Meth. in Phys.
Research Section B} {\bf 267} 244-247

\bibitem{K-K} Kirchner T and Knudsen H 2011, {\it  J. Phys. B: At. Mol.%
Opt. Phys. }{\bf 44} 122001


\bibitem{knudsen90} Knudsen H, Brun-Nielsen L, Charlton M and Poulsen M R
1990, \textit{J. Phys. B: At. Mol. Opt. Phys.} \textbf{23} 3955-3976

\bibitem{Jacobsen95} Jacobsen F M, Frandsen N P, Knudsen H, Mikkelsen U and
Schrader D M 1995, \textit{J. Phys. B: At. Mol. Opt. Phys.}
\textbf{28} 4691-4695

\bibitem{Mori94} Mori S and Sueoka O 1994, \textit{J. Phys. B: At. Mol. Opt.
Phys.} \textbf{27}, 4349-4364

\bibitem{Laricchia02} Laricchia G, Van Reeth P, Szuinska M and Moxom J 2002,
\textit{J. Phys. B: At. Mol. Opt. Phys.} \textbf{35} 2525-2540

\bibitem{Marler05} Marler J P, Sullivan J P, Surko C M 2005, \textit{Phys. Rev. A} \textbf{71}
022701

\bibitem{Charlton89} M Charltoni, L Brun-Nielsen, B I Deutch, P Hvelplund, F
M Jacobsen, H Knudsen, G Laricchiat and M R Poulsen 1989, \textit{J.
Phys. B: At. Mol. Opt. Phys.} \textbf{22} 2779-2788

\bibitem{Kara97} Kara V, Paludan K, Moxom J, Ashley P and Laricchia G 1997,
\textit{J. Phys. B: At. Mol. Opt. Phys.} \textbf{30} 3933-3949

\bibitem{Moxom96} Moxom J, Ashley P and Laricchia G 1996, \textit{Can. J.
Phys.} \textbf{74} 367

\bibitem{Moxom99} Moxom J, D. M. Schrader, G. Laricchia, Jun Xu, and L. D.
Hulett1 1999 \textit{Phys. Rev. A} \textbf{60} 2940-2943

\bibitem{bluhme99} Bluhme H, Knudsen H, Merrison J P and Nielsen K A 1999,
\textit{J. Phys. B: At. Mol. Opt. Phys.} \textbf{32} 5237

\bibitem{Moxom00} Moxom J 2000, \textit{J. Phys. B: At. Mol. Opt. Phys.}
\textbf{33} L481-L485

\bibitem{helms} Helms S, Brinkmann U, Deiwiks J, Hippler R, Schneider H,
Segers D and Paridaens J 1995,  \textit{J. Phys. B: At. Mol. Opt.
Phys.} \textbf{28} 1095

\bibitem{kruse} Kruse G, Quermann A, Raith W, Sinapius G and Weber M 1991
\textit{J. Phys. B: At. Mol. Opt. Phys.} \textbf{24} L33


\bibitem{AP} Montanari C C and Miraglia J E 2012, \textit{J. Phys. B: At.
Mol. Opt. Phys.} \textbf{45}, 105201

\bibitem{positron} Montanari C C and Miraglia J E 2015, \textit{J. Phys. B:
At. Mol. Opt. Phys.} \textbf{48}, 165203.


\bibitem{miraglia08} Miraglia J E and Gravielle M S 2008 {\it Phys. Rev.} A {\bf 78}
052705

\bibitem{FPR} Fainstein P D, Ponce V H and Rivarola R D 1988, \textit{J. Phys. B:
At. Mol. Opt. Phys.} \textbf{ 21}, 287

\bibitem{hci} Montanari C C and Miraglia J E 2015, \textit{J. Phys.: Conf. Ser.} \textbf{583} 012018

\bibitem{Santos01} Santos A C F, Melo W S, Sant 'Anna M M, Sigaud G M and
Montenegro E C 2001, \textit{Phys Rev A} \textbf{63} 062717

\bibitem{santana} Sant'Anna M M, Luna H, Santos A C F, McGrath C, Shah M B,
Cavalcanti E G, Sigaud G M and Montenegro E C 2003, \textit{Phys Rev
A} \textbf{68} 042707

\bibitem{B2} Wolff W, Luna H, Santos A C F, Montenegro E C, %
DuBois R D, Montanari C C and Miraglia J E 2011, \textit{Phys. Rev.
A} {\bf 84}, 042704

\bibitem{icpeac11} Montanari C C, Miraglia J E, Wolff W, Luna H, Santos A C
F, and Montenegro E C 2012, \textit{J. Phys.: Conf. Ser.}
\textbf{388} 012036.



\bibitem{K01} Kirchner T, Horbatsch M and L{\"u}dde H J 2001, \textit{%
Phys. Rev. A} \textbf{64}, 012711; 2002, \textit{Phys. Rev. A}
\textbf{66}, 052719

\bibitem{Kirchner05} Kirchner T, Santos A C F, Luna H, Sant'Anna M M, Melo W S,
Sigaud G M, and Montenegro E C 2005, \textit{Phys. Rev. A}
\textbf{72} 012707.

\bibitem{KirchnerAr} Schenk G and Kirchner T 2009, \textit{J. Phys. B: At.
Mol. Opt. Phys. }\textbf{42} 205202

\bibitem{Kirchner13} Schenk G, Horbatsch M and Kirchner T 2013, \textit{%
Phys. Rev. A} \textbf{88}, 012712

\bibitem{Kirchner15} Schenk G and Kirchner T 2015, \textit{%
Phys. Rev. A} \textbf{91}, 052712

\bibitem{tachino} Tachino C A, Galassi M E and Rivarola R D 2008, \textit{%
Phys. Rev. A} \textbf{77} 032714

\bibitem{tachino2} Tachino C A, Galassi M E and Rivarola R D 2009, \textit{%
Phys. Rev. A} \textbf{80} 014701.

\bibitem{sigaud13} Tavares A C , Montanari C C, Miraglia J E , and
Sigaud G M 2014, \textit{J. Phys. B: At. Mol. Opt. Phys.}
\textbf{47}, 045201.

\bibitem{archubi} Archubi C D, Montanari C C, and Miraglia J E 2007, \textit{%
J. Phys. B: At. Mol. Opt. Phys. }\textbf{40} 943


\bibitem{REG} Rapp D and Englander-Golden P 1965, \textit{J. Chem. Phys.}
\textbf{43} 1464

\bibitem{sorokinNe} Sorokin A A, Shmaenok L A, Bobashev S V, M\"{o}bus B,
and Ulm G 1998, \textit{\ Phys. Rev} A \textbf{58} 2900

\bibitem{Rudd} Rudd M E, Kim Y-K, Madison D H and Gallagher J W 1985, \textit{Rev. Modern
Phys.} \textbf{57} 965–94

}
\end{thebibliography}
\end{document}